World Scientific
www.worldscientific.com

# Self-adjoint time operator of a quantum field

Hou Y. Yau

*FDNL Research, 91 Park Manor Drive,*
*Daly City, California, USA*
*\*hyau@fdnresearch.us*

We study the properties of a quantum field with time as a dynamical variable. Temporal vibrations are introduced to restore the symmetry between time and space in a matter field. The system with vibrations of matter in time and space obeys the Klein–Gordon equation and Schrödinger equation. The energy observed is quantized under the constraint that a particle's mass is on shell. This real scalar field has the same properties of a zero-spin bosonic field. Furthermore, the internal time of this system can be represented by a self-adjoint operator without contradicting the Pauli's theorem. Neutrino can be an interesting candidate for investigating the effects of these temporal and spatial vibrations because of its extremely light weight.

## 1. Introduction

The asymmetric formulation between time and space in quantum theory has inspired the quest for a time operator. In its formulation, time is postulated as a parameter. There is nothing dynamical about time in quantum theory. On the other hand, spacetime is dynamical and weaved as unity in general relativity. There is no globally defined time in the theory. Therefore, the treatment of time is rather different in quantum theory and general relativity. This conflict has created constellation of problems when we try to reconcile the two fundamental theories from a single framework.[1,2]

The reason why time is not treated as an operator can be traced back to Pauli.[3] Based on his reasonings, widely known as Pauli's theorem, a time operator $t$ and a Hamiltonian operator $H$ should satisfy a commutation relation, $[H, t] = -i$. Since the Hamiltonian of a system is typically bounded from below or discrete, time cannot be treated as a self-adjoint operator. On the other hand, time seems to play a more dynamical role in some quantum systems, e.g. tunneling time,[4,5] decay of an unstable particle,[6] dwell time of a particle,[7,8] and others.[9–16] To avoid contradiction with Pauli's theorem, a time operator in most of the studies are formulated by invoking the positive operator valued measures (POVM).[17–21] Apart from these efforts, many propositions have also been made intending to resolve the dynamical nature of time in quantum theory. For example, Lee advocates that time can be considered as a fundamentally discrete dynamical variable in many classical and quantum models.[22,23] Greenberger infers that a few paradoxical behaviors in classical and quantum mechanics can be cured if mass is an operator and not as a simple parameter.[24] A self-adjoint time operator in the Dirac field that satisfies a commutation relation with the Hamiltonian analogous to the one between position and momentum has also been introduced by Bauer.[25]

In this paper, we show that the properties of a bosonic field can be reconciled from a field with vibrations of matter in time and space. The hypothetical temporal vibrations[26–28] are introduced to restore the symmetry between time and space in a matter field. The real scalar field describing this system with temporal and spatial vibrations obeys the Klein–Gordon equation and Schrödinger equation. Its energy must be quantized under the constraint that a particle's mass is on shell. Additionally, the internal time of this system can be represented by a self-adjoint operator. The spectrum of this operator spans the entire real line without contradicting Pauli's theorem. The concepts developed are relativistically palatable.

## 2. Vibrations of Matter in Time and Space

In classical mechanics, matter can have vibration in the spatial directions but not in the temporal direction. If nature has a preference for symmetry,[29] it is not implausible that matter can also have vibration in time. To begin, let us consider a plane wave that has vibrations of matter in time only, as observed in an inertial reference frame $O'$, i.e.

$$t'_f = t' - T_0 \sin(\omega_0 t') = t' + \text{Re}(\zeta'_t), \tag{1}$$

where

$$\zeta'_t = -i T_0 e^{-i\omega_0 t'}, \tag{2}$$

$(t', \mathbf{x}')$ are the coordinates in frame $O'$, $t'_f$ is the 'internal time' of matter, $T_0$ is a 'proper time amplitude', and $\omega_0$ is an intrinsic angular frequency of matter which we will later identify it as the frequency for mass–energy conjectured by de Broglie.[30] The meanings of $t'_f$ and $T_0$ will be explained further below.

Note that matter in this plane wave has no vibration in the spatial directions, i.e.

$$\mathbf{x}'_f = \mathbf{x}', \tag{3}$$

where $\mathbf{x}'_f$ is the coordinate of matter displaced from the equilibrium coordinate $\mathbf{x}'$ due to the vibrations in the wave. In this plane wave with vibrations of matter in proper time, the coordinates $\mathbf{x}'_f$ and $\mathbf{x}'$ are the same.

'External time' $t'$ is measured by clocks that are not coupled to the system under investigation.[31–34] These clocks are located far away at spatial infinity such that the effects from our system are negligible. We will use the external time $t'$ as a reference to measure the vibrations of time in the system. It is an independent variable in the equations of motion and a parameter used as postulated in quantum theory. There is nothing dynamical about this external time.

Time $t'_f$ is the 'internal time' of matter. It is a function of the external time $t'$ and a dynamical variable for the system. Analogous to the amplitude of a classical oscillating system with vibrations in the spatial directions, we will define the proper time displacement amplitude $T_0$ as the maximum difference between the time $t'_f$ of matter inside the wave and the external time $t'$. Subsequently, if matter inside the wave carries an internal clock, an inertial observer outside will see the matter's clock measuring a time $t'_f$, which is different from the external time $t'$. Consequently, time measured by the matter's internal clock is running at a varying rate relative to the inertial observer's clock,

$$\frac{\partial t'_f}{\partial t'} = 1 - \omega_0 T_0 \cos(\omega_0 t'), \tag{4}$$

which has an average value of 1 over time. Matter in this plane wave will appear to travel along a near time-like geodesic if the magnitude of the vibration is relatively small. On the other hand, the time displacement relative to the external time is,

$$\Delta t' = t'_f - t' = -T_0 \sin(\omega_0 t'). \tag{5}$$

This wave has the semblance of a classical oscillating system except the vibration of matter is in time and not in space. The nature of this internal time and the oscillating time displacement will be elaborated further in Sec. 3.

The background coordinates $(t', \mathbf{x}')$ of inertial frame $O'$ can be Lorentz transformed to the background coordinates $(t, \mathbf{x})$ for the flat spacetime in another frame of reference $O$, i.e.

$$t' = \gamma(t - \mathbf{v} \cdot \mathbf{x}), \tag{6}$$

$$\mathbf{x}' = \gamma(\mathbf{x} - \mathbf{v}t), \tag{7}$$

where

$$\gamma = (1 - |\mathbf{v}|^2)^{-1/2}. \tag{8}$$

We have assumed $O'$ is traveling with velocity $\mathbf{v}$ relative to $O$. Similarly, the displaced coordinates of matter $(t'_f, \mathbf{x}'_f)$ can be Lorentz transformed to the displaced coordinates of matter $(t_f, \mathbf{x}_f)$ as observed in frame $O$, i.e.

$$t_f = \gamma(t'_f + \mathbf{v} \cdot \mathbf{x}'_f), \tag{9}$$

$$\mathbf{x}_f = \gamma(\mathbf{x}'_f + \mathbf{v}t'_f). \tag{10}$$

Substitute Eqs. (1), (3), (6) and (7) into Eqs. (9) and (10), the displaced coordinates of matter $(t_f, \mathbf{x}_f)$ are

$$t_f = t + T\sin(\mathbf{k}\cdot\mathbf{x} - \omega t) = t + \text{Re}(\zeta_t), \tag{11}$$

$$\mathbf{x}_f = \mathbf{x} + \mathbf{X}\sin(\mathbf{k}\cdot\mathbf{x} - \omega t) = \mathbf{x} + \text{Re}(\zeta_\mathbf{x}), \tag{12}$$

where

$$\zeta_t = -iTe^{i(\mathbf{k}\cdot\mathbf{x} - \omega t)}, \tag{13}$$

$$\zeta_\mathbf{x} = -i\mathbf{X}e^{i(\mathbf{k}\cdot\mathbf{x} - \omega t)}, \tag{14}$$

$$\omega^2 = \omega_0^2 + |\mathbf{k}|^2, \quad T^2 = T_0^2 + |\mathbf{X}|^2, \quad \mathbf{v} = \mathbf{k}/\omega. \tag{15}$$

Amplitude $\mathbf{X}$ is the maximum displacement of matter from its equilibrium coordinate $\mathbf{x}$, and amplitude $T$ is its maximum displacement from the external time $t$. The proper time displacement $T_0$ can be seen as a Lorentz transformation of a 4-displacement vector: $(T_0, 0, 0, 0) \rightarrow (T, \mathbf{X})$.

Apart from the vibrations in time, matter in the plane wave also has vibrations in space as observed in frame $O$. At a particular instant, matter is displaced from its equilibrium coordinate $\mathbf{x}$ to another coordinate $\mathbf{x}_f$ as shown in Eq. (12). This spatial vibration of matter is the same as defined in a classical system.

We can write $\zeta_t$ and $\zeta_\mathbf{x}$ in terms of a plane wave $\zeta$ for describing the temporal and spatial vibrations of matter, i.e.

$$\zeta = \frac{a}{\omega_0^2}e^{i(\mathbf{k}\cdot\mathbf{x} - \omega t)}, \tag{16}$$

where

$$\zeta_t = \partial_0\zeta, \tag{17}$$

$$\zeta_\mathbf{x} = -\nabla\zeta, \tag{18}$$

$$a = \omega_0 T_0, \tag{19}$$

$$T = (\omega/\omega_0)T_0, \quad \mathbf{X} = (\mathbf{k}/\omega_0)T_0. \tag{20}$$

In the rest of this paper, we will consider $T$, $\mathbf{X}$ and $T_0$ as complex amplitudes.

The plane wave $\zeta$ and its complex conjugate $\zeta^*$ satisfy the wave equations:

$$\partial_u\partial^u\zeta + \omega_0^2\zeta = 0, \tag{21}$$

$$\partial_u\partial^u\zeta^* + \omega_0^2\zeta^* = 0. \tag{22}$$

The corresponding Lagrangian density for the equations of motion is

$$\mathcal{L} = K[(\partial^u\zeta^*)(\partial_u\zeta) - \omega_0^2\zeta^*\zeta], \tag{23}$$

and the Hamiltonian density is,

$$\mathcal{H} = K[(\partial_0\zeta^*)(\partial_0\zeta) + (\nabla\zeta^*)\cdot(\nabla\zeta) + \omega_0^2\zeta^*\zeta], \tag{24}$$

where $K$ is a constant of the system to be determined. Therefore, the plane wave $\zeta$ satisfies an equation of motion similar to the Klein–Gordon equation. However, we shall bear in mind that, so far, there is nothing that requires the energy in this wave to be quantized.

## 3. Proper Time Oscillator

Let us consider a plane wave from Eq. (16) that has vibrations of matter in time only ($\omega = \omega_0$ and $|\mathbf{k}| = 0$) as observed in frame $O'$,

$$\zeta' = \frac{T_0}{\omega_0} e^{-i\omega_0 t'}. \tag{25}$$

Substitute $\zeta'$ into Eq. (24), the Hamiltonian density is

$$\mathcal{H}_0 = 2K T_0^* T_0. \tag{26}$$

This result is similar to the Hamiltonian density of a harmonic oscillating system in classical mechanics, except that the vibrations are in time. Analogous to its classical counterpart, we make an ansatz,

$$K = \frac{m\omega_0^2}{2V}, \tag{27}$$

for a system that can have multiple number of point particles with mass $m$ in a cube with volume $V$. Periodic boundary conditions are imposed on the box walls.

From Eqs. (26) and (27), the energy inside volume $V$ is,

$$E = m\omega_0^2 T_0^* T_0. \tag{28}$$

The vibration in proper time is an intrinsic property of matter. Energy $E$ shall, therefore, correspond to certain energy intrinsic to matter. Since the vibration in proper time does not involve any force fields, $E$ is not energy resulting from charges. In fact, the only energy present in this system is the matter with mass $m$. Here, we will consider $E$ as the internal mass–energy resulting from the proper time vibration of matter.

The internal mass–energy of matter must be on shell. For a system with only one particle, Eq. (28) becomes

$$E = m\omega_0^2 T_0^* T_0 = m, \tag{29}$$

or from Eq. (19),

$$\omega_0^2 T_0^* T_0 = a^* a = 1. \tag{30}$$

This implies only an oscillator with proper time amplitude

$$|\mathring{T}_0| = 1/\omega_0, \tag{31}$$

can be observed.[a] Additionally, the energy $E$ from Eq. (29) is the internal mass–energy of a point mass that is at rest. A particle observed in the plane wave $\zeta'$ has oscillation in time but with no motion in the spatial directions.

Based on Eqs. (1) and (31), the internal time $\overset{\circ}{t}{}'_f$ of the particle observed is as follows:

$$\overset{\circ}{t}{}'_f(t') = t' - \frac{\sin(\omega_0 t')}{\omega_0}. \tag{32}$$

The internal time $\overset{\circ}{t}{}'_f$ is a function of the external time $t'$. It is an intrinsic dynamical property of matter with nothing to do with the relative velocity of the particle nor gravitational effects. The varying internal time rate shall have effects on the intrinsic properties of matter, e.g. decay rate of an unstable particle. In addition to the classical concepts of mass,[35,36] we suggest here a possibility that a point mass is a temporal oscillator in time with an angular frequency of $\omega_0$. In the rest of this paper, we will consider the angular frequency $\omega_0$ as the de Broglie's frequency for the mass–energy of a particle.

We have assumed matter is not traveling along a true time-like geodesic but with vibration over time. From Eq. (32), the internal time of the particle's clock passes at the rate of $\partial_0 \overset{\circ}{t}{}'_f = 1 - \cos(\omega_0 t')$ with respect to the external time. It has an average value of 1 and bounded between 0 and 2. Subsequently, the internal time of the oscillator moves only forward. It cannot go backward to the past. On the other hand, the particle will appear to travel along a time-like geodesic if the observer's clock is not sensitive enough to detect the high frequency of the oscillation. The accuracy of the measuring clock shall be restricted by the energy–time uncertainty relation.[24,37]

From Eq. (32), the time displaced from the external time $t'$ of the proper time oscillator is

$$\overset{\circ}{\Delta}t' = \overset{\circ}{t}{}'_f(t') - t' = -\frac{\sin(\omega_0 t')}{\omega_0}. \tag{33}$$

The rate of this oscillating time displacement relative to the external time is

$$\frac{\partial(\overset{\circ}{\Delta}t')}{\partial t'} = -\cos(\omega_0 t'). \tag{34}$$

The external time $t'$ is the 'equilibrium position' of this oscillating system. When the internal time is displaced from its 'equilibrium position', the system tries to return to its equilibrium. After the internal time reaches its equilibrium, the nonzero oscillating time displacement rate causes the internal time to pass over the equilibrium. This system is analogous to a classical simple harmonic oscillator except the oscillation is in time.

As shown in Eq. (29), the oscillation of matter in time can give rise to the mass–energy $E$ of a particle which can also be written in terms of $\overset{\circ}{\Delta}t'$ and $\partial(\overset{\circ}{\Delta}t')/\partial t'$, i.e.

---

[a] A ring symbol $\circ$ on the top denotes the quantity is a property of the observed particle.

$$E = m\omega_0^2(\overset{\circ}{\Delta}t')^2 + m\left[\frac{\partial(\overset{\circ}{\Delta}t')}{\partial t'}\right]^2 = m. \tag{35}$$

The internal mass–energy $E$ is the summation of two parts. The first part is the energy arising from the oscillating time displacement, $\overset{\circ}{\Delta}t'$. The second part is the energy resulting from the oscillating time displacement rate, $\partial(\overset{\circ}{\Delta}t')/\partial t'$. They are analogous to the 'potential' and 'kinetic' energy components of a classical harmonic oscillator. The total energy of this oscillator is conserved over time. There is no energy flowing in and out.

Next, let us consider a plane wave $\zeta'$ with amplitude $T_0 = 1/\omega_0$ as observed in frame $O'$, i.e.

$$\zeta' = \frac{e^{-i\omega_0 t'}}{\omega_0^2}. \tag{36}$$

The Hamiltonian density of this plane wave from Eq. (24) is $\mathcal{H}_0 = m/V$. If we probe only a part of the system with volume $V_1$ $(< V)$, the energy observable is $E_1 = mV_1/V$. It is only a fraction of a particle's mass–energy. However, if a particle is point-like and its mass–energy is on shell, how can we explain the presence of a fraction of a particle in volume $V_1$?

In a quantum wave, a probability density can be defined for the observation of a particle at a particular location. To explain the presence of a fraction of the particle's mass–energy, we can treat the plane wave $\zeta'$ as a probabilistic wave. In other words, there is only a probability of observing a particle in the probed volume $V_1$. Based on the Hamiltonian density $\mathcal{H}_0$, we can define a probability density for the plane wave $\zeta'$, i.e.

$$\rho = \mathcal{H}_0/m = 1/V. \tag{37}$$

The probability for observing a particle in volume $V_1$ is $p_1 = V_1/V$. After making many measurements with similar experimental setup, the average mass–energy observable in volume $V_1$ is $E_1 = mV_1/V$.

## 4. Moving Particle with Vibration in Time

In the normalized plane wave $\zeta'$ from Eq. (36), one particle with rest mass $m = \omega_0$ can be observed when we probe the system as a whole. As discussed, the particle observed has oscillation in proper time with an amplitude $|\overset{\circ}{T}_0| = 1/\omega_0$. From Eq. (16), the Lorentz transformation of the normalized plane wave $\zeta'$ to another frame $O$ is

$$\zeta = \frac{e^{i(\mathbf{k}\cdot\mathbf{x}-\omega t)}}{\omega_0^2}. \tag{38}$$

Frame $O'$ is assumed to be traveling at a velocity $\mathbf{v} = \mathbf{k}/\omega$ relative to frame $O$. The particle observed in frame $O$ shall be traveling with an average velocity $\mathbf{v}$, but with oscillations in time and space after the Lorentz transformation of the proper time oscillation; its energy is $E = \omega$. However, when we examine the Hamiltonian density of the normalized plane wave $\zeta$ obtained from Eq. (24)

$$\mathcal{H} = \omega/V', \tag{39}$$

it is equivalent to one particle with energy $\omega$ in a volume $V' = V\alpha^2$, where $\alpha = \sqrt{\omega_0/\omega}$. Since the system we are studying has a volume $V$ and not $V'$, it is necessary to include the normalization factor $\alpha$ with the plane waves when they are used in the superposition for a more general application.

In the rest of this paper, it is more convenient to adopt another plane wave $\tilde{\zeta}$ for our analysis, i.e.

$$\tilde{\zeta} = \alpha\zeta = \frac{T_0}{\sqrt{\omega\omega_0}} e^{i(\mathbf{k}\cdot\mathbf{x} - \omega t)}. \tag{40}$$

Substitute $\tilde{\zeta}$ into Eq. (24), the Hamiltonian density is

$$\tilde{\mathcal{H}} = \frac{m\omega\omega_0 T_0^* T_0}{V}. \tag{41}$$

Based on Eqs. (11), (12), (17) and (18), the temporal and spatial vibrations in plane wave $\tilde{\zeta}$ are

$$\tilde{t}_f = t + \text{Re}(\tilde{\zeta}_t), \tag{42}$$

$$\tilde{\mathbf{x}}_f = \mathbf{x} + \text{Re}(\tilde{\zeta}_\mathbf{x}), \tag{43}$$

where

$$\tilde{\zeta}_t = \partial_0 \tilde{\zeta} = -i\tilde{T} e^{i(\mathbf{k}\cdot\mathbf{x} - \omega t)}, \tag{44}$$

$$\tilde{\zeta}_\mathbf{x} = -\nabla\tilde{\zeta} = -i\tilde{\mathbf{X}} e^{i(\mathbf{k}\cdot\mathbf{x} - \omega t)}, \tag{45}$$

$$\tilde{T} = \alpha T = T_0 \sqrt{\frac{\omega}{\omega_0}}, \quad \tilde{\mathbf{X}} = \alpha\mathbf{X} = \frac{T_0 \mathbf{k}}{\sqrt{\omega_0\omega}}. \tag{46}$$

For a normalized plane wave with $T_0 = 1/\omega_0$, the Hamiltonian density from Eq. (41) is $\tilde{\mathcal{H}} = \omega/V$. This is equivalent to one particle with energy $\omega$ in a volume $V$. Since the system that we are investigating has a constant volume $V$, we shall utilize $\tilde{\zeta}$ when the superposition principle is applied.

Assuming a particle in the plane wave $\tilde{\zeta}$ is first observed at origin of the $\mathbf{x}$ coordinates at $t = 0$. After Lorentz transforming the proper time oscillator observed in plane wave $\zeta'$ and including the normalization factor $\alpha$ with the amplitudes of oscillation, the internal time of the particle following the path $\mathbf{x} = \mathbf{v}t$ as observed in $\tilde{\zeta}$ is

$$\overset{\circ}{t}_f = t - \overset{\circ}{T}\sin(\omega_p t), \tag{47}$$

where

$$\mathring{T} = \sqrt{\frac{\omega}{\omega_0^3}}, \tag{48}$$

$$\omega_p = \frac{\omega_0^2}{\omega}. \tag{49}$$

The internal time rate relative to the external time is

$$\frac{\partial \mathring{t}_f}{\partial t} = 1 - \sqrt{\frac{\omega_0}{\omega}} \cos(\omega_p t). \tag{50}$$

Apart from the oscillation in time, the particle observed also has oscillation in the spatial directions. Its trajectory is,

$$\mathring{\mathbf{x}}_f = \mathbf{v}t - \mathring{\mathbf{X}} \sin(\omega_p t), \tag{51}$$

where

$$\mathring{\mathbf{X}} = \frac{\mathbf{k}}{\sqrt{\omega_0^3 \omega}}. \tag{52}$$

The observed velocity with oscillation is,

$$\frac{\partial \mathring{\mathbf{x}}_f}{\partial t} = \mathbf{v}\left[1 - \sqrt{\frac{\omega_0}{\omega}} \cos(\omega_p t)\right]. \tag{53}$$

We shall note that as $|\mathbf{v}| \rightarrow 1$, the magnitude of the amplitudes approach infinity, i.e. $|\mathring{T}| \rightarrow \infty$ and $|\mathring{\mathbf{X}}| \rightarrow \infty$. On the other hand, $\omega_p$ is the angular frequency of a moving particle. It is not the angular frequency $\omega$ of the plane wave. As $|\mathbf{v}| \rightarrow 1$, the angular frequency $\omega_p$ slows down and approaches zero, $\omega_p \rightarrow 0$. Subsequently, a particle traveling at a higher speed will have a lower frequency and larger amplitudes of oscillation.

As we have discussed in the previous section, the plane wave $\zeta'$ with proper time vibrations shall be treated as a probabilistic wave. Similarly, the same concept shall be applied to plane wave $\tilde{\zeta}$. The Hamiltonian density from Eq. (41) can be written in terms of $\tilde{\zeta}$, i.e.

$$\tilde{\mathcal{H}} = \frac{\omega_0^3 \omega^2}{V} \tilde{\zeta}^* \tilde{\zeta}. \tag{54}$$

Since the energy of an observed particle in plane wave $\tilde{\zeta}$ is $E = \omega$, we can define a probability density of observing a particle as,

$$\rho = \frac{\tilde{\mathcal{H}}}{\omega} = \frac{\omega_0^3 \omega}{V} \tilde{\zeta}^* \tilde{\zeta}. \tag{55}$$

There is only a probability of observing a particle with energy $\omega$ at a particular location in plane wave $\tilde{\zeta}$.

## 5. Wave Function

Based on Eq. (55), we can define a function $\psi$ in terms of the plane wave $\tilde{\zeta}$ in the nonrelativistic limit

$$\psi = \frac{\omega_0 T_0}{\sqrt{V}} e^{i(\mathbf{k}\cdot\mathbf{x} - \bar{\omega}t + \chi)} \approx \left[\sqrt{\frac{\omega_0^3 \omega}{V}} e^{i(\omega_0 t + \chi)}\right] \tilde{\zeta}, \tag{56}$$

where

$$\bar{\omega} = \mathbf{k}\cdot\mathbf{k}/(2\omega_0) \approx \omega - \omega_0, \tag{57}$$

and $e^{i\chi}$ is an arbitrary phase factor. From Eqs. (19), (55) and (56), the square modulus of the function $\psi$,

$$\rho = \psi^*\psi = \frac{\omega_0^2 T_0^* T_0}{V} = \frac{a^* a}{V} = \frac{n}{V}, \tag{58}$$

is a probability density; $n = a^* a$ is the number of particle that can be observed in volume $V$. As discussed in the previous section, there is only a probability to observe a particle at a location. Function $\psi$ has the basic properties of a wave function in quantum mechanics.

Applying the superposition principle, we can write

$$\psi(\mathbf{x}, t) = \sum_{\mathbf{k}} \frac{a_{\mathbf{k}}}{\sqrt{V}} e^{i(\mathbf{k}\cdot\mathbf{x} - \bar{\omega}t + \chi)}, \tag{59}$$

where

$$a_{\mathbf{k}} = \omega_0 T_{0\mathbf{k}}. \tag{60}$$

Boundary condition is imposed such that $\psi(\mathbf{x}, t)$ vanishes at the box walls. This wave function $\psi(\mathbf{x}, t)$ is a solution of the linear and homogeneous Schrödinger equation, i.e. $i\dot{\psi}(\mathbf{x}, t) = -(2m)^{-1}\nabla^2\psi(\mathbf{x}, t)$. The probability amplitude $a_{\mathbf{k}}$ of the wave function can be expressed in terms of the proper time vibration amplitude $T_{0\mathbf{k}}$ from Eq. (60). The probability density of observing a particle is the square modulus of $\psi(\mathbf{x}, t)$. As we have illustrated, the properties of the quantum mechanical wave can be reconciled from the system with vibrations of matter in time and space.

In quantum mechanics, it is commonly believed that a matter wave can only have a probabilistic interpretation because of the unobservable phase for the wave function $\psi$.[37] As we shall note, the introduction of the arbitrary phase factor $e^{i\chi}$ does not change the probability density calculated in Eq. (58). The wave function with the arbitrary phase factor still satisfies the Schrödinger equation. Therefore, the plane wave $\tilde{\zeta}$ and the wave function $\psi$ can have an arbitrary phase difference but this difference will not alter the results obtained in quantum mechanics. As demonstrated in quantum mechanics, the theory developed with wave functions $\psi$ is invariant under global phase transformation but the relative phase factors are physical.

Although the overall phase of the wave function $\psi$ can be unobservable, it serves only as a mathematical tool for describing an underlying wave with vibrations of matter in time and space. Here, we demonstrate a possibility that the matter wave can have a physical interpretation other than the probabilistic one despite the overall phase of the wave function is unobservable.

## 6. Bosonic Field

We can obtain a real scalar field by the superposition of the plane waves $\tilde{\zeta}$ and their conjugates $\tilde{\zeta}^*$, i.e.

$$\zeta(\mathbf{x}) = \frac{1}{\sqrt{2}} \sum_{\mathbf{k}} [\tilde{\zeta}_{\mathbf{k}}(\mathbf{x}) + \tilde{\zeta}_{\mathbf{k}}^*(\mathbf{x})]$$
$$= \sum_{\mathbf{k}} (2\omega\omega_0)^{-1/2} [T_{0\mathbf{k}} e^{-i\mathbf{k}\cdot\mathbf{x}} + T_{0\mathbf{k}}^* e^{i\mathbf{k}\cdot\mathbf{x}}]. \tag{61}$$

To adopt the same convention in quantum field theory, we will define

$$\varphi(\mathbf{x}) = \sum_{\mathbf{k}} (2\omega V)^{-1/2} [\omega_0 T_{0\mathbf{k}} e^{-i\mathbf{k}\cdot\mathbf{x}} + \omega_0 T_{0\mathbf{k}}^* e^{i\mathbf{k}\cdot\mathbf{x}}]$$
$$= \zeta(\mathbf{x}) \sqrt{\frac{\omega_0^3}{V}}, \tag{62}$$

which satisfies the Klein–Gordon equation. This real scalar field is subject to the boundary condition that $\varphi(\mathbf{x})$ vanishes at the box walls.

As discussed in the previous sections, the system with matter vibrating in time and space shall be treated as a probabilistic wave. By allowing matter to vibrate in both the temporal and spatial directions, we can reconcile the properties of a quantum wave in the nonrelativistic limit. The system considered is explained by discrete particles and not a spatially continuous field. As developed in quantum field theory, the transition of a classical field to a quantum field can be done via canonical quantization. Here, we can also adopt the same concept to obtain a quantum field from the system that has vibrations of mater in time and space. In other words, the fields $\varphi(\mathbf{x})$ and $\zeta(\mathbf{x})$ are to be promoted to operators. Since the quantization of a real scalar field is a familiar process, we will only outline some of the key results involving the temporal vibrations that are not part of the standard quantum theory.

We can show that $\varphi(\mathbf{x})$ is the same bosonic field in quantum theory after we rewrite Eq. (62) in terms of the creation operator,

$$a_{\mathbf{k}}^\dagger = \omega_0 T_{0\mathbf{k}}^\dagger, \tag{63}$$

and the annihilation operator,

$$a_{\mathbf{k}} = \omega_0 T_{0\mathbf{k}}, \tag{64}$$

*H. Y. Yau*

such that

$$\varphi(\mathbf{x}) = \sum_{\mathbf{k}} (2\omega V)^{-1/2} [a_{\mathbf{k}} e^{-i\mathbf{k}\cdot\mathbf{x}} + a_{\mathbf{k}}^{\dagger} e^{i\mathbf{k}\cdot\mathbf{x}}]. \tag{65}$$

The operators $a_{\mathbf{k}}$, $a_{\mathbf{k}}^{\dagger}$, $T_{0\mathbf{k}}$ and $T_{0\mathbf{k}}^{\dagger}$ shall satisfy the commutation relations,

$$[a_{\mathbf{k}}, a_{\mathbf{k}'}^{\dagger}] = \delta_{\mathbf{k}\mathbf{k}'}, \tag{66}$$

$$[a_{\mathbf{k}}, a_{\mathbf{k}'}] = [a_{\mathbf{k}}^{\dagger}, a_{\mathbf{k}'}^{\dagger}] = 0, \tag{67}$$

$$[T_{0\mathbf{k}}, T_{0\mathbf{k}'}^{\dagger}] = \frac{\delta_{\mathbf{k}\mathbf{k}'}}{\omega_0^2}, \tag{68}$$

$$[T_{0\mathbf{k}}, T_{0\mathbf{k}'}] = [T_{0\mathbf{k}}^{\dagger}, T_{0\mathbf{k}'}^{\dagger}] = 0. \tag{69}$$

In fact, by expressing the creation and annihilation operators in terms of $T_{0\mathbf{k}}$ and $T_{0\mathbf{k}}^{\dagger}$, we can rewrite other operators in quantum theory using the temporal and spatial vibrations field. For example, the particle number operator is,

$$N_{\mathbf{k}} = a_{\mathbf{k}}^{\dagger} a_{\mathbf{k}} = \omega_0^2 T_{0\mathbf{k}}^{\dagger} T_{0\mathbf{k}}. \tag{70}$$

The Hamiltonian of the system is,

$$H = \sum_{\mathbf{k}} \omega \left( a_{\mathbf{k}}^{\dagger} a_{\mathbf{k}} + \frac{1}{2} \right) = \sum_{\mathbf{k}} \omega \left( \omega_0^2 T_{0\mathbf{k}}^{\dagger} T_{0\mathbf{k}} + \frac{1}{2} \right), \tag{71}$$

which corresponds to an infinite sum of normal mode oscillator excitation. Since the conversion of other operators are straightforward, we will not repeat them in here. Note that normal ordering shall be taken between $T_{0\mathbf{k}}$ and $T_{0\mathbf{k}}^{\dagger}$.

As we have demonstrated, the formulation of a bosonic field can be expressed in terms of the vibrations of matter in time and space. As a quantum field, the energy in the system considered shall be quantized; the oscillators in proper time are the field quanta. The results obtained here have the familiar properties of a zero-spin bosonic field except they are obtained from a field that has vibrations of matter in time and space.

## 7. Internal Time Operator

After quantization, the field $\zeta(\mathbf{x})$ from Eq. (61) can be rewritten in terms of the temporal vibration amplitude operators $\tilde{T}_{\mathbf{k}}$ and their Hermitian conjugates $\tilde{T}_{\mathbf{k}}^{\dagger}$, i.e.

$$\zeta(\mathbf{x}) = \frac{1}{\sqrt{2}} \sum_{\mathbf{k}} \left[ \frac{\tilde{T}_{\mathbf{k}}}{\omega} e^{-i\mathbf{k}\cdot\mathbf{x}} + \frac{\tilde{T}_{\mathbf{k}}^{\dagger}}{\omega} e^{i\mathbf{k}\cdot\mathbf{x}} \right], \tag{72}$$

where

$$\tilde{T}_{\mathbf{k}} = \sqrt{\frac{\omega}{\omega_0}} T_{0\mathbf{k}} = \sqrt{\frac{\omega}{\omega_0^3}} a_{\mathbf{k}}, \tag{73}$$

$$\tilde{T}_{\mathbf{k}}^{\dagger} = \sqrt{\frac{\omega}{\omega_0}} T_{0\mathbf{k}}^{\dagger} = \sqrt{\frac{\omega}{\omega_0^3}} a_{\mathbf{k}}^{\dagger}. \tag{74}$$

In addition, $\tilde{T}_{\mathbf{k}}^{\dagger}$ and $\tilde{T}_{\mathbf{k}}$ satisfy the commutation relations,

$$[\tilde{T}_{\mathbf{k}}, \tilde{T}_{\mathbf{k}'}^{\dagger}] = \frac{\omega}{\omega_0^3} \delta_{\mathbf{k}\mathbf{k}'}, \tag{75}$$

$$[\tilde{T}_{\mathbf{k}}, \tilde{T}_{\mathbf{k}'}] = [\tilde{T}_{\mathbf{k}}^{\dagger}, \tilde{T}_{\mathbf{k}'}^{\dagger}] = 0. \tag{76}$$

From Eq. (17), the field $\zeta_t(\mathbf{x})$ describing the temporal vibrations in $\zeta(\mathbf{x})$ is,

$$\zeta_t(\mathbf{x}) = \partial_0 \zeta(\mathbf{x}) = \sum_{\mathbf{k}} \frac{-i}{\sqrt{2}} [\tilde{T}_{\mathbf{k}} e^{-i\mathbf{k}\cdot\mathbf{x}} - \tilde{T}_{\mathbf{k}}^{\dagger} e^{i\mathbf{k}\cdot\mathbf{x}}]. \tag{77}$$

In fact, $\zeta_t(\mathbf{x})$ can be expressed in terms of the conjugate momentum of $\zeta(\mathbf{x})$.

From Eq. (23), the Lagrangian density for $\zeta(\mathbf{x})$ is

$$\mathcal{L} = \frac{\bar{\rho}_m \omega_0^2}{2} [(\partial_0 \zeta)^2 - (\nabla \zeta)^2 - \omega_0^2 \zeta^2], \tag{78}$$

where

$$\bar{\rho}_m = \frac{\omega_0}{V} \tag{79}$$

is a mass density constant of the system. Hence, the conjugate momentum of $\zeta(\mathbf{x})$ is

$$\begin{aligned} \eta(\mathbf{x}) &= \frac{\partial \mathcal{L}}{\partial[\partial_0 \zeta(\mathbf{x})]} = \bar{\rho}_m \omega_0^2 \zeta_t(\mathbf{x}) \\ &= \frac{-i\bar{\rho}_m \omega_0^2}{\sqrt{2}} \sum_{\mathbf{k}} [\tilde{T}_{\mathbf{k}} e^{-i\mathbf{k}\cdot\mathbf{x}} - \tilde{T}_{\mathbf{k}}^{\dagger} e^{i\mathbf{k}\cdot\mathbf{x}}]. \end{aligned} \tag{80}$$

Therefore, both $\zeta_t(\mathbf{x})$ and $\eta(\mathbf{x})$ can be used to describe the temporal vibrations in the real scalar field.

The conjugate pair $\zeta(\mathbf{x})$ and $\eta(\mathbf{x})$ are Hermitians and satisfy the equal-time commutation relations:

$$[\zeta(t, \mathbf{x}), \eta(t, \mathbf{x}')] = i\delta(\mathbf{x} - \mathbf{x}'), \tag{81}$$

$$[\zeta(t, \mathbf{x}), \zeta(t, \mathbf{x}')] = [\eta(t, \mathbf{x}), \eta(t, \mathbf{x}')] = 0. \tag{82}$$

Similarly,

$$[\zeta(t, \mathbf{x}), \zeta_t(t, \mathbf{x}')] = (\bar{\rho}_m \omega_0^2)^{-1} \delta(\mathbf{x} - \mathbf{x}'), \tag{83}$$

$$[\zeta_t(t, \mathbf{x}), \zeta_t(t, \mathbf{x}')] = 0. \tag{84}$$

As we have learned from quantum theory, the real scalar field $\varphi(\mathbf{x})$ and its conjugate momentum are self-adjoint operators. From Eq. (62), we show that $\varphi(\mathbf{x})$ can be expressed in terms of $\zeta(\mathbf{x})$. Therefore, it is not surprising that $\zeta(\mathbf{x})$, $\zeta_t(\mathbf{x})$ and $\eta(\mathbf{x})$ are also self-adjoint operators. Additionally, the conjugate of $\eta(\mathbf{x})$ is $\zeta(\mathbf{x})$ and not the Hamiltonian as shown in Eq. (80). Therefore, there is no commutation relation with the semi-bounded Hamiltonian that restricts the spectrum of the temporal vibration operator [using $\zeta_t(\mathbf{x})$ or $\eta(\mathbf{x})$] to be bounded.

Based on Eq. (11), the internal time in the real scalar field at a particular time $t$ is

$$t_f(t, \mathbf{x}) = t + \zeta_t(t, \mathbf{x}), \tag{85}$$

where $\zeta_t(t, \mathbf{x})$ is real as shown in Eq. (77). The internal time $t_f(t, \mathbf{x})$ is the summation of the temporal vibration $\zeta_t(t, \mathbf{x})$ and the external time $t$. Since the temporal vibration $\zeta_t(t, \mathbf{x})$ is a self-adjoint operator and the external time $t$ is a parameter, the internal time $t_f(t, \mathbf{x})$ is also a self-adjoint operator. The symmetry between time and space is restored in the system with both the internal time and position of matter can be treated as self-adjoint operators.

## 8. Conclusions and Discussions

Unlike the asymmetrical formulation of time in the classical and quantum theory, time and space have a more symmetrical treatment in the system considered. Matter in this system can vibrate not only in the spatial directions but also in the temporal direction. The symmetry between time and space is restored in the system with both the internal time and position of matter can be treated as self-adjoint operators. Here, we demonstrate a possibility that time can have a more dynamical role in the quantum field. By allowing matter to vibrate in both time and space, we have reconciled the properties of a bosonic field.

The reason why the temporal vibration $\zeta_t$ can be treated as a self-adjoint operator is that we are considering an oscillating system. An oscillator with temporal vibration can have displacement either in the positive or negative direction relative to the external time $t$. Therefore, its spectrum can span the whole real line. As we have pointed out in the previous section, the internal time $t_f$ can also be treated as a self-adjoint operator since it is the summation of a parameter $t$ and a self-adjoint operator $\zeta_t$. This internal time operator does not form a conjugate pair with the Hamiltonian of the system. Therefore, the restriction imposed by Pauli's theorem does not apply in this case. The inclusion of time oscillation in a matter field is a new approach that we have adopted in this paper.

The temporal oscillation can theoretically affect the rate of change for the intrinsic properties of a particle (e.g. decay rate of an unstable particle). Additionally, a particle has oscillation in the spatial directions as it propagates through space. To examine the magnitude of these oscillations, let us consider a neutrino, which is the lightest known elementary particle. Assuming the mass of a neutrino[b] is $m = 2\,\text{eV}$ ($\omega_0 = 3.04 \times 10^{15}\,\text{s}^{-1}$ and $\overset{\circ}{T}_0 = 1/\omega_0 = 3.29 \times 10^{-16}\,\text{s}$), the amplitudes and frequency of a moving neutrino can be obtained from Eqs. (48), (49) and (52), e.g.

---

[b]At present, the absolute mass scale of neutrino has not yet been determined. The assumed mass $m = 2\,\text{eV}$ is the upper limit of the electron-neutrino mass determined by direct measurements as shown in Refs. 38 and 39.

$$E = 1\,\mathrm{Gev} \Rightarrow \overset{\circ}{T} = 7.4 \times 10^{-12}\,\mathrm{s},$$
$$|\overset{\circ}{\mathbf{X}}| = 0.22\,\mathrm{cm}, \quad \omega_p = 6.1 \times 10^6\,\mathrm{s}^{-1}. \tag{86}$$

The effects of the temporal and spatial oscillations shall be considered when we are taking measurements of a neutrino. As discussed in Sec. 4, a particle traveling at a higher speed will have a lower frequency and larger amplitudes of oscillation. Consequently, it will be easier to detect the hypothetical effects of the additional oscillations at a higher speed. For instance, the spatial oscillation of a particle is along the direction of propagation. A neutrino will propagate with oscillating motions along its path. Therefore, two neutrinos with the same initial velocity can reach a target at slightly different times depending on the relative phase of their oscillations. In theory, this deviation can be observable by repeated measurements of the neutrinos' arrival times at a detector. Because of their extremely light weight, neutrinos can be projected to a very high speed which can amplify the oscillations for measurements. Neutrino can be an interesting candidate for investigating the effects of these temporal and spatial oscillations.